\documentclass[twocolumn,showpacs,prl]{revtex4}

\usepackage{amsmath,amssymb}
\usepackage{graphicx}
\usepackage{verbatim}

\draft
\begin{document}

\title{Quantum Anomalous Hall Effect with Cold Atoms Trapped in a
Square Lattice}

\author{Xiong-Jun Liu$^a$, Xin Liu$^a$, Congjun Wu$^b$, and Jairo Sinova$^a$}
\affiliation{a. Department of Physics, Texas
A\&M University, College Station, Texas 77843-4242, USA\\
b. Department of Physics, University of California, San Diego,
California 92093, USA}

\begin{abstract}
We propose an experimental scheme to realize the quantum anomalous
Hall effect in an anisotropic square optical lattice which can be
generated from available experimental set-ups of double-well
lattices with minor modifications. A periodic gauge potential
induced by atom-light interaction is introduced to give a Peierls
phase for the nearest-neighbor site hopping. The quantized
anomalous Hall conductivity is investigated by calculating the
Chern number as well as the chiral gapless edge states of our
system. Furthermore, we show in detail the feasability for its
experimental detection through light Bragg scattering of the edge
and bulk states with which one can determine the topological phase
transition from usual insulating phase to quantum anomalous Hall
phase.
\end{abstract}
\date{\today }
\maketitle

\section{Introduction}

Twenty years ago, Haldane proposed a 
toy model in the honeycomb lattice to illustrate the quantum
anomalous Hall effect (QAHE) \cite{Haldane}, in which a complex
second-nearest-neighbor hopping term $t_2e^{i\phi}$ drives the
system into the topologically insulating state. Different from the
conventional quantum Hall effect (QHE) \cite{QHE1}, Landau levels
are not necessary for QAHE, while in both the QHE and QAHE
systems, the time-reversal symmetry (TRS) is broken. The quantized
Hall conductivity can be explained with Laughlin's gauge
invariance argument \cite{Laughlin} and by Halperin's edge state
picture \cite{halperin}, which has a deep topological reason  as
the first Chern class of U(1) principal fiber bundle on a torus
\cite{TKNN}. Considering the topological nontriviality and the
absence of magnetic field of the QAHE, realizing experimentally
this new state of matter in its cleanest form is of fundamental
importance in the study of new materials such as topological
insulators.

Unfortunately, Haldane's model cannot be realized in the recently
discovered graphene system, since the required staggered magnetic
flux in the model is extremely hard to achieve. A recent proposal
predicts the QAHE in the Hg$_{1-x}$Mn$_x$Te quantum wells
\cite{QAHE1} by doping Mn atoms in the quantum spin Hall system of
the HgTe quantum well to break TRS \cite{zhang,zhang1,experiment}.
QAHE is reachable within a time range much smaller than the
relaxation time of Mn spin polarization which is about $10^{-4}$s,
while so far the experimental study of this effect is not
available. On the other hand, the technology of ultracold atoms in
optical lattices allows for a controllable fashion unique access
to the study of condensed matter physics. An artificial version of
the staggered magnetic field (Berry curvature) with hexagonal
symmetry is considered to obtain Haldane's model for cold atoms
trapped in a honeycomb optical lattice \cite{zhu1}. While a
periodic Berry curvature can be readily obtained by coupling
atomic internal states to standing waves of laser fields
\cite{atomgauge1,atomgauge4,atomgauge6,experiment1,hou}, the
experimental realization of the staggered magnetic field with
hexagonal symmetry remains a challenge. Furthermore, Wu shows QAHE
can be reached with the $p$-orbital band in the honeycomb optical
lattice by applying a technique developed in S. Chu's group
\cite{gemelke2007} to rotate each lattice site around its own
center \cite{wu}.

In this work, we propose a distinct realization of QAHE in a
two-dimensional (2D) anisotropic square optical lattice, which can
be realized based on the double-well experiments performed at NIST
\cite{NIST}, superposed with a periodic gauge potential which is
also experimentally accessible. The experimental detection of
quantum anomalous Hall (QAH) states is also proposed and
investigated in detail through light Bragg scattering.

\section{The model}

We consider an anisotropic 2D square optical lattice depicted in
Fig. 1 (a) filled with fermions (e.g. $^6$Li, $^{40}$K), whose
optical potential is expressed as $V_{latt}(\bold r)
=-V_{0}(\cos^2k_0x+\cos^2k_0y) - V_0\cos^2 [\frac{k_0}{2}(x+y)
+\frac{\pi}{2} ]$. This potential can be generated from the
available experimental set-up of the double-well lattice
illustrated in the Fig 1b of Ref. \onlinecite{NIST} by placing an
additional polarizing beam splitter (PBS) before the mirror
reflector $M_3$ and suitably adjusting the phase difference among
different optical paths. The first term of the potential is from
the light component with the in-plane ($x$-$y$) polarization which
is deflected by the PBS and then reflected back by $M_3$, while
the second term is from the light component with the out-of-plane
($z$)-polarization which passes the PBS without reflection. No
interference exists between these two components. This optical
potential has a structure of two sublattices $A$ and $B$. The
potential minimum $V_A$ at site $A$ is higher than $V_B$ at site
$B$  as $V_A-V_B=V_0$. The anisotropic potential at site A has
different frequencies along the directions of $\hat
e_{1,2}=\frac{1}{\sqrt 2} (\hat e_x \pm \hat e_y)$ as
$\omega_1^A=(V_{0}k_0^2)^{1/2}/m^{1/2}$ and $\omega_2^A= \sqrt 2
\omega_1^A$, and those at site $B$ are $\omega_1^B=\sqrt 3
\omega_1^A$ and $\omega_2^B=\omega_2^A=\sqrt 2\omega_1^A$,
respectively. The local orbital is described by 2D harmonic
oscillator eigenstate $\psi_{n_1n_2}^{A,B}$ with the eigenvalues
\begin{eqnarray}\label{eqn:eigenvalue1}
E_{n_1,n_2}^{A}&=&(n_1^{A}+\frac{1}{2})\hbar\omega_1^{A}+(n_2^{A}+\frac{1}{2})\hbar\omega_2^{A},\nonumber\\
E_{n_1,n_2}^{B}&=&(n_1^{B}+\frac{1}{2})\hbar\omega_1^{B}+(n_2^{B}+\frac{1}{2})\hbar\omega_2^{B}-V_0.
\end{eqnarray}
Below we shall consider $V_0$ taking the value of
$V_0=(3\sqrt{3}-1) ^2E_r/2-4M$ with $M$ satisfying $M\ll E_r$ and
$E_r=\hbar^2k_0^2/2m$ the recoil energy. In this case, the
$s$-orbital at the $B$-sites is the lowest one, while the
$s$-orbital at the $A$-sites is nearly degenerate with the
$p$-orbital at the $B$-sites along the $\hat e_1$ direction with
the energy difference of $E^B_{1,0}-E^A_{00}\approx 2M$.
Specifically, if $M=0$ we have $E^B_{1,0}-E^A_{00}=0$. For
convenience, we denote such two nearly degenerate states by
$\psi_a=\psi_{0,0}^{A}$ and $\psi_b=\psi_{1,0}^B$, respectively,
which consist of a pseudospin-$1/2$ subspace.
We shall focus on the hybridized bands between $\psi_a$ and
$\psi_b$ intermediated by the intersite hopping, but neglect the
hybridization between $\psi_{a,b}$ and the lowest $s$-orbital in
the B sites since its onsite energy is far separated from those of
$\psi_{a,b}$ in the case of large trapping frequencies for the
lattice \cite{RMPcoldatom}.
\begin{figure}[ht]
\includegraphics[width=0.9\columnwidth]{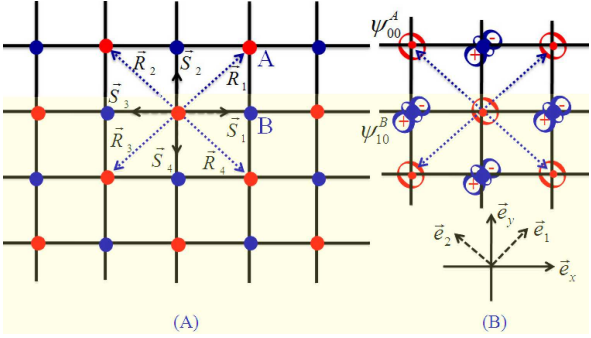}
\caption{(Color online) (a) Fermions trapped in a 2D anisotropic
square optical lattice. Due to the different on-site trapping
frequencies, the square lattice is divided into two sublattices
$A$ and $B$; (b) The local orbitals are in the $\vec e_1$ and
$\vec e_2$ directions.} \label{fig1}
\end{figure}

To break TR symmetry, we introduce a periodic
adiabatic gauge potential in the simple form $\bold A(\bold
r)=\hbar A_0\sin(k_0(y-x))\vec e_y$ with $A_0$ a constant, which
can be generated by coupling atoms to two opposite-travelling
standing-wave laser beams with Rabi-frequencies
$\Omega_1=\Omega_0\sin\bigr(k_0(y-x)/2+\pi/4\bigr)e^{-iA_0y},
\Omega_2=\Omega_0\cos\bigr(k_0(y-x)/2+\pi/4\bigr)e^{iA_0y}$
\cite{atomgauge4,atomgauge6,experiment1}. Note the gauge field can
lead to an additional contribution $|\bold A|^2/2m$
without the square lattice symmetry 
\cite{honeycomb}, which, however, will not distort the present
square lattice. This is because, first of all, the term $|\bold
A|^2/2m$ is zero at all square lattice sites; secondly, the
contributed potential of this term is along the $\vec e_2$
direction and leads to the same correction to $\omega_2^{A,B}$.
Therefore we still have $\omega_2^A=\omega_2^B$ and the properties
of $\psi_{a,b}$ are unchanged. This is a key difference from the
situation in honeycomb lattice system, where a gauge potential
with hexagonal symmetry is required to avoid the distortion of
original honeycomb lattice \cite{zhu1,honeycomb}. The simple form
of gauge potential indicates here a feasible scheme in the
experimental realization.

The periodic gauge potential gives rise to a Peierls phase for the
hopping coefficients obtained by $\exp(i\phi^{\bold r_i}_{\bold
L})=\exp(i\int_{\bold r_i}^{\bold r_i+\bold L}\bold A\cdot d\bold
r/\hbar)$, where the integral is along the hopping path from the
site $\bold r_i$ to site $\bold r_i+\bold L$. Taking into account
the hopping between both the nearest and second-nearest neighbor
sites, we obtain the Hamiltonian in the tight-binding form:
$H=H_1+H_2+H_z$, with $H_z=M\sum_{\bold r_i}[C_b^{\dag}(\bold
r_i)C_b(\bold r_i)-C_a^{\dag}(\bold r_i)C_a(\bold r_i)]$, and
\begin{eqnarray}\label{eqn:tightbinding1}
H_1&=&-\bigr[\sum_{\bold r_i}\sum_{j=1,2}t_{ab}e^{i\phi^{\bold
r_i}_{\bold S_j}}\hat C_a^{\dag}(\bold r_i)\hat C_b(\bold
r_i+\bold
S_j)+h.c.\bigr]\nonumber\\
&&+\bigr[\sum_{\bold r_i}\sum_{j=3,4}t_{ab}e^{i\phi^{\bold
r_i}_{\bold S_j}}\hat C_a^{\dag}(\bold r_i)\hat C_b(\bold
r_i+\bold
S_j)+h.c.\bigr],\nonumber\\
H_2&=&-\sum_{\bold
r_i}\sum_{\mu=a,b}\biggr[\sum_{j=1,3}t_{\mu1}e^{i\phi^{\bold
r_i}_{\bold R_j}}\hat C_\mu^{\dag}(\bold r_i)\hat C_\mu(\bold
r_i+\bold
R_j)\nonumber\\
&&+\sum_{j=2,4}t_{\mu2}e^{(i\phi^{\bold r_i}_{\bold R_j}}\hat
C_\mu^{\dag}(\bold r_i)\hat C_\mu(\bold r_i+\bold R_j)\biggr],
\end{eqnarray}
where $\hat C_\mu(\bold r_i)$ is the annihilation operator on site
$\bold r_i$ in sublattices $A$ (for $\mu=a$) and $B$ (for
$\mu=b$), the vectors $\bold S_1(-\bold S_3)=(a,0), \bold
S_2(-\bold S_4)=(0,a), \bold R_1(-\bold R_3)=(a,a)$ and $\bold
R_2(-\bold R_4)=(-a,a)$, with $a=\pi/k_0$ the lattice constant. It
is easy to check that the Peierls phase $\phi_{R_j}^{\bold
r_i}=\phi_{S_{1,3}}^{\bold r_i}=0$, while $\phi_{S_{2,4}}^{\bold
r_i}=\phi_0$ (or $\phi_{S_{2,4}}^{\bold r_i}=-\phi_0$) with
$\phi_0=\sqrt{2}A_0/k_0$ when the $\bold r_i$ site belongs to
sublattice $A$ (or sublattice $B$). From the symmetry of the wave
functions $\psi_{a,b}$ we know the hopping coefficients satisfy
$t_{a1}, t_{a2}, t_{b2}, t_{ab}>0; t_{b1}<0$, $t_{a1}\neq t_{a2}$
and $t_{b1}\neq t_{b2}$ (Fig. 1(b)). Besides, since the hopping
constants decay exponentially with distance, $t_{ab}$ is typically
several times bigger in magnitude than $t_{aj}$ and $t_{bj}$
$(j=1,2)$. Nevertheless, such differences will not affect the
topological phase transition. Finally, noting $\psi_{a,b}$ have
the same spatial distribution in the $\vec e_2$ direction, we can
expect that $t_{a2}\approx t_{b2}$ or $|t_{a2}-t_{b2}|\ll
t_{a1}-t_{b1}$.

It is convenient to transform the tight-binding Hamiltonian into
momentum space, say, let $\hat C_{a,b}(\bold
r_j)=\frac{1}{\sqrt{N}}\sum_{\bold k}e^{i\bold k\cdot\bold
r_j}\hat C_{a,b}(\bold k)$ and we obtain Hamiltonian in the matrix
form $H=\sum_{\bold k}\hat {\mathcal C}^{\dag}(\bold
k)\mathcal{H}(\bold k)\hat {\mathcal C}(\bold k)$ with $\hat
{\mathcal C}(\bold k)=(\hat C_a(\bold k), \hat C_b(\bold k))^T$
and (neglecting the constant terms)
\begin{eqnarray}\label{eqn:Hamiltonian1}
\mathcal{H}(\bold k)=\lambda_x(\bold k)\sigma_x+\lambda_y(\bold
k)\sigma_y+\lambda_z(\bold k)\sigma_z.
\end{eqnarray}
Here $\lambda_x=2t_{ab}\sin\phi_0\sin(k_ya)$,
$\lambda_y=2t_{ab}(\sin(k_xa)+\cos\phi_0\sin(k_ya))$ and
$\lambda_z=-M-\Delta_0 \cos(k_xa)\cos(k_ya)
-2\tilde{t}_1\sin(k_xa)\sin(k_ya)$ with
$\tilde{t}_1=(t_{a1}-t_{b1}+t_{b2}-t_{a2})/2, \Delta_0
=t_{a1}-t_{b1}+t_{a2}-t_{b2}$. As long as $\phi_0\neq n\pi$, the
TR symmetry of the system is broken. This Hamiltonian leads to two
energy bands with the spectra given by ${\cal
E}_\pm=\pm\sqrt{\sum_i\lambda_i^2(\bold k)}$. One can check when
$\phi_0\neq n\pi$, the band gap is opened if at the two
independent Dirac points $\lambda_z(\bold K_c)\neq0$, with $\bold
K_C=(0,0), (0,\pi)$. Therefore when $M\neq\pm\Delta_0$
the system is gapped in the bulk.

\begin{figure}[ht]
\includegraphics[width=0.9\columnwidth]{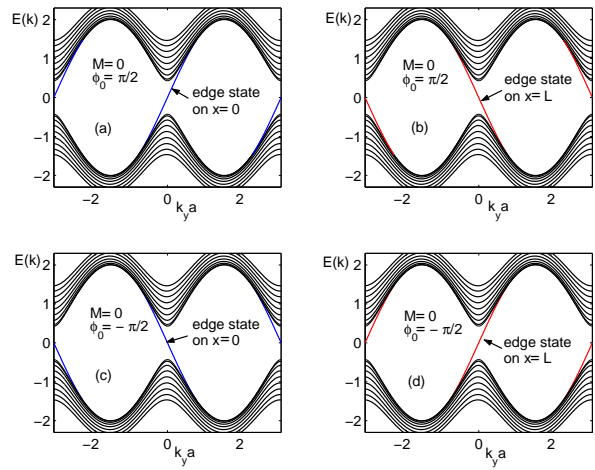}
\caption{(Color online) Gapped bulk states (black lines) and
gapless edge states (blue and red lines) on the boundaries $x=0$
((a)(c)) and $x=L$ ((b)(d)). Parameters in (a-b) are
$M=0,\phi_0=\pi/2$, and in (c-d) are $M=0,\phi_0=-\pi/2$. The
chirality of edge states indicates the Chern number $C_1=+1$ for
$\phi_0=\pi/2$ and $C_1=-1$ for $\phi_0=-\pi/2$.} \label{edge1}
\end{figure}

When the Fermi energy 
is inside the band gap, the longitudinal conductivity is zero. The
anomalous Hall conductivity (AHC) can be calculated by
\begin{eqnarray}\label{eqn:AHC1}
\sigma_{xy}^{H}=C_1/\hbar,
\end{eqnarray}
where $C_1=(4\pi)^{-1}\int_{FBZ}dk_xdk_y\bold
n\cdot\partial_{k_x}\bold n\times\partial_{k_y}\bold n$ with
$\bold n(\bold
k)=(\lambda_x,\lambda_y,\lambda_z)/|\vec\lambda(\bold k)|$, is the
first Chern number, a quantized topological invariant defined on
the first Brillouin zone (FBZ). Actually one can construct a
mapping degree between FBZ torus and spherical surface $S^2$, $F:
S^1\times S^1\mapsto S^2$, which gives the Chern number
$C_1=m\in\mathbb{Z}$, with $m$ the times the mapping covers the
$S^2$ surface. By a straightforward calculation we can show in the
case $-\Delta_0<M<\Delta_0$ and thus the effective masses of the
Dirac Hamiltonian around two independent Dirac points are opposite
in the sign, the Chern number $C_1=+1$ when $0<\phi_0<\pi$ and
$C_1=+1$ when $-\pi<\phi_0<0$. In all other cases we have $C_1=0$.
The quantized AHC can support topological stable gapless edge
states on the boundaries of the system. To study the edge modes,
we consider a hard-wall boundary \cite{hardwall} along $x$ axis at
$x=0$ and $x=L$. The momentum $k_x$ is no longer a good quantum
number, and we shall transform the terms with $k_x$ in the
Hamiltonian back to position space. For convenience we envisage
first the case $\phi_0=\pi/2$ and obtain that
($H=\sum_{k_y,x_i}\mathcal{H}(k_y, x_i)$)
\begin{eqnarray}\label{eqn:matrixform11}
H&=&\sum_{k_y, x_i}\bigr[\mathcal{M}\hat C_{k_y, x_i}^\dag\hat
C_{k_y, x_i}+\mathcal{A}\hat C_{k_y,x_i}^\dag\hat
C_{k_y,x_i+1}\nonumber\\
&&+\mathcal{A}^\dag\hat C_{k_y,x_i+1}^\dag\hat C_{k_y,x_i}\bigr],
\end{eqnarray}
where
\begin{eqnarray}\label{eqn:matrixform12}
\mathcal{M}=2t_{ab}\sin k_ya\sigma_x-M\sigma_z,
\end{eqnarray}
and
\begin{eqnarray}\label{eqn:matrixform13}
\mathcal{A}&=&-it_{ab}\sigma_y-\sum_{j=1,2}\frac{t_{aj}-t_{bj}}{2}\cos
k_ya\sigma_z\nonumber\\
&&+\frac{i}{2}(t_{a1}-t_{b1}+t_{b2}-t_{b1})\sin k_ya\sigma_z.
\end{eqnarray}
Note the edge modes are generally exponentially localized on the
boundary \cite{edge1}. Denoting by $\Psi^{+}(k_y, x_i)$ and
$\Psi^{-}(k_y, x_i)$ the edge states on the boundaries $x=0$ and
$x=L$, respectively, one can verify that they take the following
form
\begin{eqnarray}\label{eqn:exactedge1}
\Psi^{\pm}(k_y, x_i)=\frac{u_{k_y}(y)}{\sqrt{{\cal
N}_{\pm}}}\bigr[(\xi^{(\pm)}_1)^{x_i/a}-(\xi^{(\pm)}_2)^{x_i/a}\bigr]\psi^\pm,
\end{eqnarray}
where the complex variables $\xi_{1,2}^\pm$ are obtained by
$\xi^-_{1,2}=\bigr[M\pm\sqrt{M^2+4(t_{ab}-i\tilde{t}_1\sin
k_ya)^2-\Delta_0^2\cos^2k_ya}\bigr]\bigr(2t_{ab}-2\tilde{t}_2\cos
k_ya-i2\tilde{t}_1\sin k_ya)\bigr)^{-1}$ and
$\xi^+_{1,2}=1/\xi^-_{1,2}$, the spinor $\psi^\pm$ satisfies
$\sigma_x\psi^\pm=\pm\psi^\pm$, $u_{k_y}(y)$ is the Bloch wave
function along $y$ axis and ${\cal N}_{\pm}$ the normalization
factor. The chirality of the edge modes can be found from their
spectra ${\cal E}^\pm_{k_y}=\pm2t_{ab}\sin k_ya$, respectively
associated with group velocities $v_F\approx\pm2at_{ab}/\hbar$
around Dirac point. Besides, the exponential decay of the edge
states on boundaries requires that $|\xi^+_{1,2}|<1$ and
$\xi^-_{1,2}>1$ \cite{edge1}. At the Dirac point $k_y=0$, one can
check such inequalities lead to $-\Delta_0<M<\Delta_0$, which is
consistent with the condition for nonzero Chern numbers obtained
before. The Fig. \ref{edge1} depicts the energy spectra in
different situations.

\section{Detection of the topological phase transition}

Next we proceed to study the detection of the edge and bulk states with
light Bragg scattering, with which one can detect the topological
phase transition. In the Bragg spectroscopy, we shine two lasers
on the lattice system, with the wave vectors $\bold k_1,\bold k_2$
and frequencies $\omega_{1,2}=k_{1,2}c$, respectively (Fig. 3(a)).
Note only the momentum $k_y$ is still a good quantum number, we
let $\bold q=\bold k_1-\bold k_2=q\hat e_y$ and denote
$\omega=\omega_1-\omega_2$. The atom-light interacting Hamiltonian
reads
\begin{eqnarray}\label{eqn:structure0}
H_{int}=\sum_{k_{1y},k_{2y}}\Omega e^{i\bold q\cdot \bold r}\hat
C_\mu^\dag(k_{y}+\bold q)\hat C_\nu(k_{y})+h.c.,
\end{eqnarray}
where $\Omega$ is an effective Rabi-frequency of the two-photon
process in light Bragg scattering, and the indices $\mu,\nu$ may
represent the edge or bulk states. The light Bragg scattering
directly measures the dynamical structure  factor\cite{Bragg}:
\begin{eqnarray}\label{eqn:structure1}
S(\bold
q,\omega)&=&\sum_{k_{y_1},k_{y_2}}(1-f(E^{(f)}_{k_{y_2}}))f(E^{(i)}_{k_{y_1}})\times\nonumber\\
&\times&|\langle\Psi_{k_{y_2}}^{(f)}|H_{int}|\Psi_{k_{y_1}}^{(i)}\rangle|^2
\delta(\hbar\omega-E^{(f)}_{k_{y_2}}+E^{(i)}_{k_{y_1}})
\end{eqnarray}
with $|\Psi_{k_{y_1}}^{(i)}\rangle$
($|\Psi_{k_{y_2}}^{(f)}\rangle$) initial (final) atomic state
before (after) scattering and $f(E)$ the Fermi distribution
function. For the topological insulating phase, e.g. when
$\phi_0=\pi/2$ and $M=0$, since the edge states are localized on
the boundaries, we may consider two basic situations for the Bragg
scattering, say, first we shine the two lasers on one boundary (on
$x=0$ or $x=L$) of the system; secondly we shine them on the whole
lattice system including both boundaries. For the former case,
only edge states on the boundary shined with lasers can be
scattered. When $\omega<\Delta_0$, the initial edge states below
Fermi energy will be scattered to edge states above Fermi energy,
while for $\omega>\Delta_0$, part of the initial edge states can
be pumped to upper band bulk states after scattering. Note that
the edge state is an exponential decaying function in the $x$
direction, with decaying property dependent on the momentum
$k_{y_1}$, while the bulk states are standing waves along $x$
axis. Therefore, the scattering process with an edge state pumped
to bulk states actually includes many channels characterized by
different values of $k_{x_2}$ of the final bulk states, and the
effective Rabi-frequency for such scattering processes is
generally a function of $k_{y_1}, k_{x_2}$, denoted by
$\tilde{\Omega}(k_{y_1},k_{x_2})$. Nevertheless, in the practical
case, we require $\omega$ is slightly above $\Delta_0$, and only
the states with momenta near zero need to be considered. In this
way we can expect $\tilde{\Omega}$ does not change considerably
from $\tilde{\Omega}_0\equiv\tilde{\Omega}(0,0)$, and can be
expanded around this value. Specifically, in continuum limit one
finds the exponential decay property of the edge state (localized
on $x=0$) given in Eq. (\ref{eqn:exactedge1}) can be approximated
as $\Psi^+\sim e^{-q_0x}$ with $q_0=\frac{\Delta_0}{2at_{ab}}$,
with which we obtain the effective Rabi-frequency
$\tilde{\Omega}(k_{x},k_{y})\approx\tilde{\Omega}_0/(q_0^2+k_x^2)$.
Bearing these results in mind, we can verify for $|q|\geq q_0$,
the dynamical structure takes the following general form
\begin{eqnarray}\label{eqn:structure6}
S(\bold
q,\omega)&\approx&(|q|+k_{y_0})\Omega^2\delta(\omega\mp v_Fq)\nonumber\\
&+&\frac{\Delta_0\tilde{\Omega}^2_0}{2t_{ab}^2a^2}\frac{1+3\hbar\tilde{\omega}/(2q_0at_{ab})}
{[1+\hbar\tilde{\omega}/(q_0at_{ab})]^2}\times\nonumber\\
&\times&\biggr[\frac{\pi}{2}+\sin^{-1}\frac{\alpha^{1/2}|q-q_0|}{\sqrt{\hbar\tilde{\omega}}}\biggr]\Theta(\omega-\omega_c),
\end{eqnarray}
where $\tilde{\omega}=\omega-\Delta_0/2\hbar-2|q|at_{ab}/\hbar,
\omega_c=(\Delta_0/2+2|q|at_{ab})/\hbar,
\alpha=2\Delta_0^{-1}t_{ab}^2a^2$, and the step function
$\Theta(x)=1 (0)$ for $x>0$ ($x<0$). For $q<q_0$, the $S(\bold
q,\omega)$ is obtained in the same form, only with $\omega_c$
changed to be $\omega_c=(\Delta_0+2a^2t_{ab}^2q^2/\Delta_0)/\hbar$
in Eq. (\ref{eqn:structure6}). The first term in $S(\bold
q,\omega)$ is contributed by the scattering processes with edge
states pumped to edge states on the boundary $x=0$ (for ``$-$") or
$x=L$ (for ``$+$"). The peaks at $\omega=\pm qv_F$ obtained by
this term reflects the chirality of the edge modes (Fig. 3(b-c),
the blue solid line). From the second term in $S(\bold q,\omega)$
we see the scattering processes with initial edge states pumped to
the upper band bulk states approximately give rise to a finite
contribution (Fig. 3(b-c), the red dashed line). For the latter
situation the lasers are shined on the whole lattice, the cross
scattering process with edge states on one boundary scattered to
edge states on another boundary will happen. We can show that the
cross scattering leads to an additional contribution
$\frac{\Omega^2}{\hbar v_F}\Theta(\omega\mp v_Fq)$ to the
dynamical structure (Fig. 3(c), the blue dotted line).
\begin{figure}[ht]
\includegraphics[width=1.0\columnwidth]{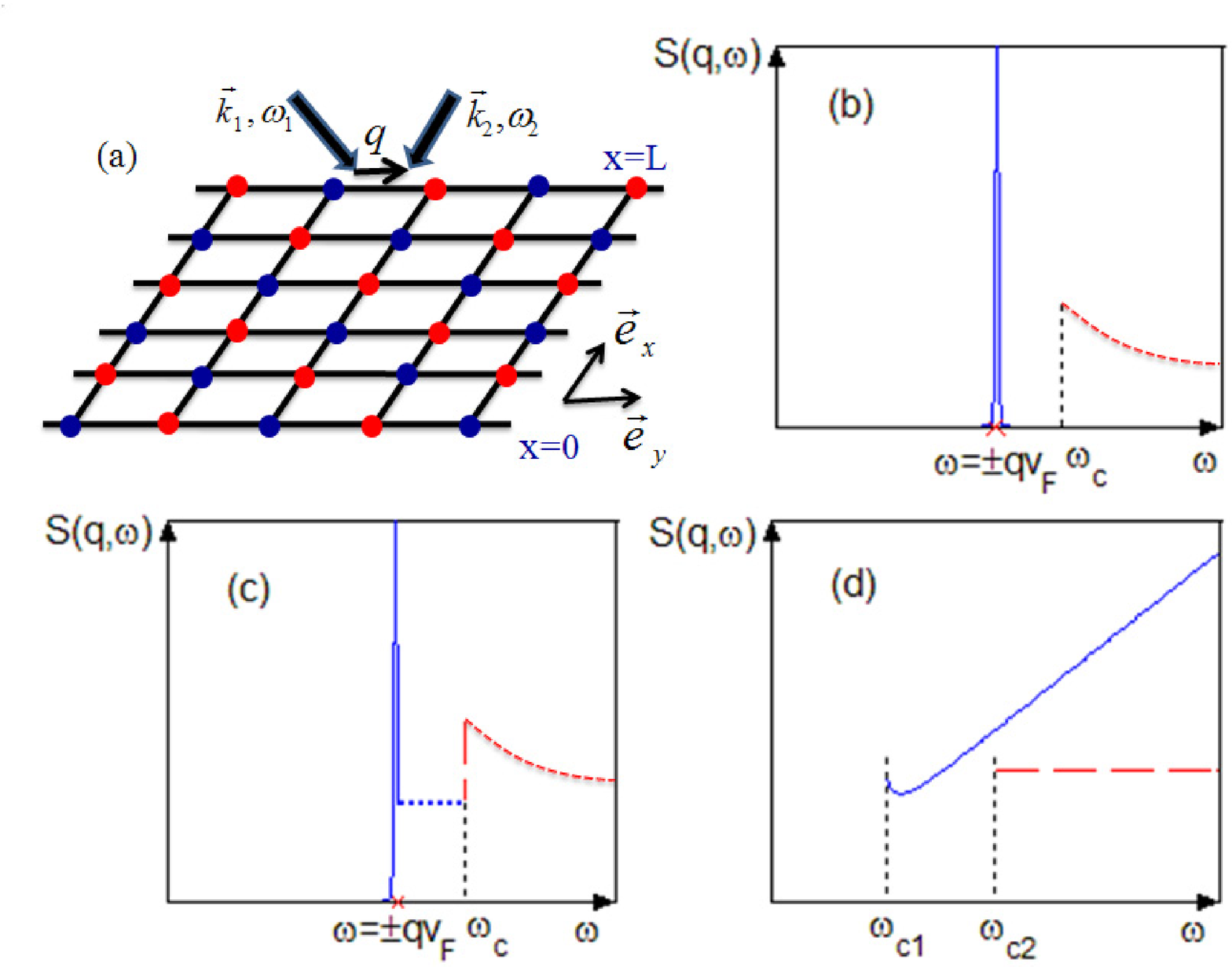}
\caption{(Color online) Schematic of light Bragg Scattering (a)
and dynamical structure for the cases that the system is in the
topological phase (b-c), and in the insulating phase (d).}
\label{Bragg}
\end{figure}

Finally, when $|M|\geq\Delta_0$, the Chern number becomes zero and
in this case no edge state can survive on the boundaries.
Specifically, for the case $M=-\Delta_0$ (and $\phi_0=\pi/2$), the
bulk gap is closed ($\Delta=0$ at the Fermi point $\bold k=0$) and
the particles can be described as massless Fermions \cite{duan},
while for $M=-2\Delta_0$, the bulk gap is given by
$\Delta=2\Delta_0$. We respectively obtain the dynamical structure
for the two cases $S(\bold
q,\omega)=\frac{\pi\Omega^2}{16a^2t_{ab}^2}\frac{\hbar^2
\omega^2-2a^2t_{ab}^2q^2}{\sqrt{\hbar^2\omega^2-4a^2t_{ab}^2q^2}}\Theta(\omega-\omega_{c_1})$
and $S(\bold
q,\omega)=\frac{\pi\Delta_0\Omega^2}{4a^2t_{ab}^2}\Theta(\omega-\omega_{c_2})$,
with $\omega_{c_1}=2at_{ab}q/\hbar$ and
$\omega_{c_2}=(2\Delta_0+q^2a^2t_{ab}^2/\Delta_0)/\hbar$. The
results are plotted with blue solid line (for $\Delta=0$) and red
dashed line (for $\Delta=2\Delta_0$) in Fig. 3(d). Based on the
results of light Bragg scattering in different situations, we
clearly see the Bragg spectroscopy provides a direct way to
observe the edge states and bulk states.

\section{conclusion}

In conclusion, we proposed a novel scheme to realize the quantum
anomalous Hall effect in an anisotropic square optical lattice
based on the experiment set-up of the double-well lattice at NIST.
We have studied in detail the experimental detection of the edge
and bulk states through light Bragg scattering, with which one can
determine the topological phase transition from usual insulating
phase to quantum anomalous Hall phase.

This work was supported by ONR under Grant No. ONR-N000140610122,
by NSF under Grant No. DMR-0547875, and by SWAN-NRI.  Jairo Sinova
is a Cottrell Scholar of the Research Corporation.
C.W. is supported by NSF-DMR0804775.


\noindent


\begin{thebibliography}{99}

\bibitem{Haldane} F. D. M. Haldane, Phys. Rev. Lett. {\bf 61}, 2015 (1988).

\bibitem{QHE1} K.V. Klitzing, G. Dorda, and M. Pepper,
Phys. Rev. Lett. {\bf 45}, 494 (1980).

\bibitem{Laughlin} R. B. Laughlin, Phys. Rev. B {\bf 23}, 5632 (1981).

\bibitem{halperin} B. I. Halperin, Phys. Rev. B {\bf 25}, 2185 (1982).

\bibitem{TKNN} D.J. Thouless, M. Kohmoto, M.P. Nightingale, and M. den Nijs, Phys.
Rev. Lett., 49, 405 (1982).

\bibitem{QAHE1} C. -X. Liu, X. -L. Qi, X. Dai, Z. Fang and S. -C. Zhang,
Phys. Rev. Lett. {\bf 101}, 146802 (2008).

\bibitem{zhang} X. -L. Qi,  Y.S. Wu, and S.C. Zhang,
Phys. Rev. B \textbf{74}, 045125 (2006).

\bibitem{zhang1} B.A. Bernevig, Taylor L. Hughes and S.-C. Zhang, Science,
{\bf 314}, 1757, (2006).

\bibitem{experiment} M. K\"{o}nig {\it et al.}, Science {\bf 318}, 766
(2007).

\bibitem{zhu1} L. B. Shao, S. -L. Zhu, L. Sheng, D.Y. Xing and Z. D.Wang,
Phys. Rev. Lett. {\bf 101}, 246810 (2008).

\bibitem{atomgauge1} J. Ruseckas, G. Juzeliunas, P. Ohberg, and M. Fleischhauer,
Phys. Rev. Lett. {\bf 95}, 010404 (2005).

\bibitem{atomgauge4} S.-L. Zhu, H. Fu, C.-J. Wu, S.-C. Zhang and L.-M. Duan,
Phys. Rev. Lett. {\bf 97}, 240401 (2006).

\bibitem{atomgauge6} X. -J. Liu, Mario F. Borunda, Xin Liu, and Jairo Sinova,
Phys. Rev. Lett. \textbf{102}, 046402 (2009).

\bibitem{experiment1} Y. -J. Lin, R. L. Compton, A. R. Perry, W. D. Phillips, J. V. Porto, and I. B. Spielman,
Phys. Rev. Lett. {\bf 102}, 130401 (2009).

\bibitem{hou} J. -M. Hou, Wen-Xing Yang, and Xiong-Jun Liu,
Phys. Rev. A {\bf 79}, 043621 (2009).

\bibitem{wu} Congjun Wu, Phys. Rev. Lett. \textbf{101}, 186807
(2008).

\bibitem{gemelke2007} N. Gemelke, Ph.D. thesis, Stanford University, 2007.

\bibitem{NIST} J. Sebby-Strabley {\it et al.}, Phys. Rev. A 73, 033605 (2006);
Phys. Rev. Lett. 98, 200405 (2007).

\bibitem{honeycomb} Tudor D. Stanescu {\it et al.}, Phys. Rev. A \textbf{79}, 053639 (2009).

\bibitem{RMPcoldatom} I. Bloch, Jean Dalibard, Wilhelm Zwerger,
Rev. Mod. Phys. \textbf{80}, 885 (2008).

\bibitem{hardwall} T. P. Meyrath, F. Schreck, J. L. Hanssen, C.-S. Chuu, and M. G. Raizen,
Phys. Rev. A {\bf 71}, 041604(R) (2005).

\bibitem{edge1} M. K\"{o}nig, Hartmut Buhmann, Laurens W. Molenkamp, Taylor Hughes, Chao-Xing Liu, Xiao-Liang Qi1, and Shou-Cheng Zhang
J. Phys. Soc. Jpn. {\bf 77}, 031007 (2008).

\bibitem{Bragg} D.M. Stamper-Kurn {\it et al.}, Phys. Rev. Lett. \textbf{83}, 2876
(1999).

\bibitem{duan} S. -L. Zhu, Baigeng Wang, and L.-M. Duan,
Phys. Rev. Lett. {\bf 98}, 260402 (2007).

\end{thebibliography}
\end{document}